# Energy Efficient Laser-Based Optical Wireless Communication Networks

Walter Zibusiso Ncube[1], Ahmad Adnan Qidan[2], Taisir El-Gorashi[2] and Jaafar M. H. Elmirghani[2]
[1]*School of Electronic and Electrical Engineering University of Leeds, Leeds, United Kingdom*
[2]*Department of Engineering, Kings' College London, London, United Kingdom*
el18wzn@leeds.ac.uk, ahmad.qidan@kcl.ac.uk, taisir.elgorashi@kcl.ac.uk, jaafar.elmirghani@kcl.ac.uk

**ABSTRACT**
Rising data demands are a growing concern globally. The task at hand is to evolve current communication networks to support enhanced data rates while maintaining low latency, energy consumption and costs. To meet the above challenge, Optical Wireless Communication (OWC) technology is proposed as a solution to complement traditional Radio Frequency (RF) based communication systems. Recently, Vertical Cavity Surface Emitting Lasers (VCSELS) have been considered for data transmission in OWC indoor environments due to their ability to transmit power through narrow, near-circular beams to receivers. In this paper, we study the energy efficiency of a VCSEL-based OWC system in an indoor environment and compare it to that of a Light Emitting Diode (LED) based system. The main findings show that the VCSEL system performs better and has higher energy efficiency.
**Keywords**: Optical Wireless Communication Networks, Vertical Cavity Surface Emitting Lasers, Zero Forcing, Interference Management, Consumption Factor, Energy Efficiency.

## 1. INTRODUCTION

The growing uptake of Internet of Things (IoT) devices in both commercial and non-commercial sectors is a rising concern globally. This growth results from and spurs the continuous development of technologies that underpin IoT. This proliferation of data-hungry IoT devices is expected to present an unprecedented demand for broadband data with which the RF spectrum cannot keep up [1]–[7]. OWC promises to provide high aggregate data rates with the potential to surpass the Gb/s barrier through the use of wide, unregulated, and free infrared and visible light spectrums [8]–[10]. However, current OWC deployments use LED devices that have limited capacity due to the low modulation bandwidth of LED devices [11], [7].

The use of laser diodes is considered due to their ability to send information at high speeds [12], [13]. Specifically, Vertical Cavity Surface Emitting Lasers (VCSELS) have received a lot of attention as transmitters for short-distance communication due to several advantages, including low production costs, high bandwidth and high-quality beams [14], [15]. They are also considered energy-efficient transmitters and have demonstrated a Consumption Factor (CF) of up to 6 Gb/mJ whilst operating at 1060 nm and achieving over 10 Gb/s [16], [17].

Previous literature analysed only a restricted set of solutions and ignores comparison to existing LED OWC deployments. In [18], a link capable of 40-56 Gb/s with an energy-to-data ratio (EDR) of 4.5 pJ/bit is demonstrated. In [17] and [19] a CF for VCSEL systems of 2 Gb/mJ and 3.75Gb/mJ, respectively, is reported. In our previous work [20], we studied the energy efficiency of a VCSEL-based optical wireless communication network under different laser beam waists to find the effective laser beam size that results in throughput enhancement. Despite the promising results of these studies, there is no comparison with the energy efficiency of current indoor deployments.

In this paper, the energy efficiency of a VCSEL-based OWC system is studied and compared against that of an LED system. It is important to note that both VCSEL and LED systems use equivalent electrical models to allow a fair comparison.

The rest of this paper is organised as follows: Section 2 gives the proposed system model. The results are given and discussed in Section 3, and the conclusions are presented in Section 4.

## 2. SYSTEM MODEL

We consider a downlink OWC system in an indoor environment. On the ceiling, multiple optical access points (APs) are deployed to service multiple users distributed on the receiving plane, which is 1 m above the ground, as shown in Figure 1. Each optical AP is a VCSEL micro-lens array, and each user is equipped with a wide Field of View (FOV) optical detector that points toward the ceiling. We aim to determine the energy efficiency of the VCSEL-based OWC system and compare its performance against a traditional LED system.

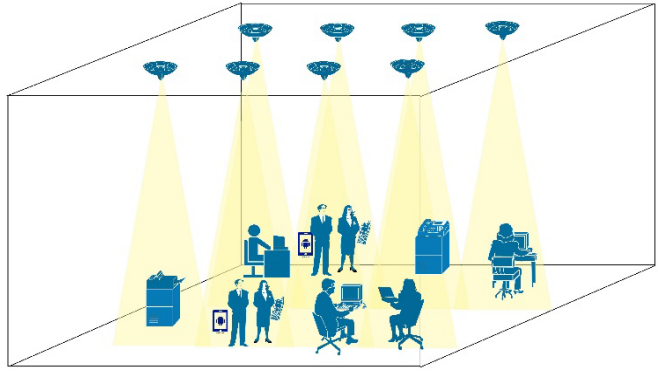

*Figure 1: System Model.*

To manage multi-user interference (MUI), we use Zero Forcing (ZF), a transmit precoding scheme that tries to remove MUI at a given user by creating a pre-coder that is orthogonal to the channel matrices of other interfering users and thereby maximising the network's overall sum rates, and consequently, achieving high energy efficiency. The system is simulated in MATLAB for the indoor environment described above. The simulation was carried out using ray tracing algorithm as in [21]. Conventional On-Off Keying (OOK) modulation is assumed, and only Line-Of-Sight (LOS) components are considered as they contribute the most significant portion of the power received. The system performance is determined by the SINR, which is calculated as follows:

$$\gamma_u^a = \frac{signal}{noise} = \frac{P_{r,u}^{a,}}{\delta_t}. \qquad (1)$$

The signal power received by user *u* from access point *a* is calculated as:

$$P_{r,u}^{a,} = R P_u^{a,} H_u^{a,} G_i^a \ (i \neq u), \qquad (2)$$

where $P_u^{a,}$ is the optical power allocated from AP, *a*, to user *u*. $H_u^{a,}$ is the DC channel gain between AP *a* and user *u*, $G_i^a$ is the ZF precoder, and R is the responsivity of the photodetector in (A/W). The total noise penalty is calculated as:

$$\delta_T = \sqrt{\left(\delta_{Rx}^2 + \delta_{Th}^2 + \delta_{RIN}^2 + \delta_u^{b^2}\right)}, \qquad (3)$$

where $\delta_u^b$ is the e background light shot noise, $\delta_{Th}$ is the thermal noise, $\delta_{RIN}$ is the Relative Intensity Noise (RIN), and $\delta_{Rx}$ is the preamplifier noise.

## 3. RESULTS

In this section, we investigate the performance of our VCSEL system, considering the power spent by the transmitters in sending data to the users. For comparison, we design a traditional LED-based VLC network with eight access points serving an equivalent number of users as the VCSEL system. The power spent by the LED was obtained from data sheets [22] and given in Table 1. Note that the number of access points for VLC is chosen to ensure coverage and illumination in the same room size. It is worth mentioning that the locations of the APs and the users are the same for both the LED and the VCSEL systems. Table 1 gives the parameters used in this work.

In Figure 2, the achievable data rates of the systems are depicted against the number of users. The results show that the VCSEL system performs better than the LED system. This is because of the large bandwidth of the VCSELs and the large amount of collected optical power in the VCSEL system. In other words, the large amount of collected power from VCSELs is relatively higher than LEDs because VCSELs have very narrow beams that can be collimated. When this beam lands on the communication floor (CF), it can have a small spot. On the other hand, LEDs have a wider beam (resulting in a larger spot on the CF) and can sometimes cover the whole floor, which might result in a waste of power. In our previous work [20], it was shown that the output power of VCSELs can be enhanced by the addition of a microlens, which results in each user receiving higher radiated power. However, eye safety needs to be considered.

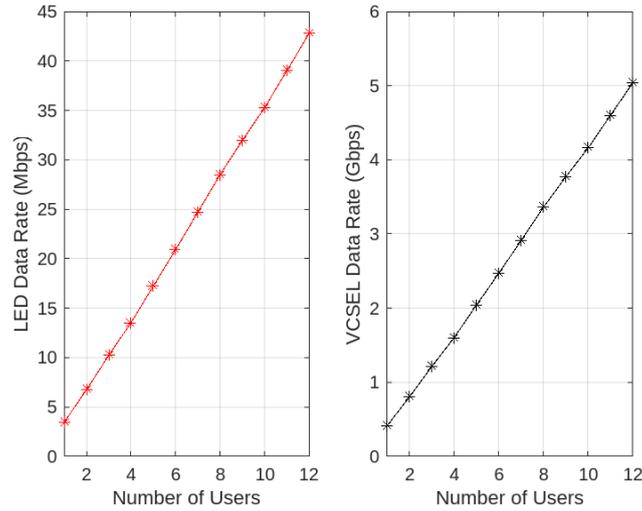
*Figure 2: Achievable Data Rate vs Number of Users.*

*Table 1: System Parameters.*

| Parameters | Configurations | Parameters | Configurations |
|---|---|---|---|
| Beam waist, $W_0$ | 5 $\mu$m | VCSEL Bandwidth | 1.5 GHz |
| Pitch Distance | 10 $\mu$m | Laser noise PSD | -155 dB/Hz |
| Link Distance | 3m | Lens refractive index | 1.5 |
| VCSELs per transmitter | 25 | Load resistance | 50 Ohms |
| Lens focal length | 0.127 mm | TIA noise figure | 5 dB |
| VCSEL to lens | 0.133 mm | FEC limit | $10^{-3}$ |
| VCSEL Array Spot size on CF | 1.5$m^2$ | Number of LED units | 8 |
| VCSEL wavelength | 1550 nm | Number of LEDs per unit | 4 |
| Energy Spent by VCSEL | 50mW | Energy Spent by LED | 3W |

Figure 3 shows the energy efficiency of the systems. Interestingly, the energy efficiency increases with the number of users due to the low consumed power for 12 users compared to the high sum rate. However, it is expected that at a high number of users, the energy efficiency might start to decrease considerably. This is because as the number of users increases, the power received by each user reduces, leading to resource contention. This results in decreased energy efficiency as the transmit power may need to be increased to maintain the same Quality of Service (QoS). The higher energy efficiency of the VCSEL system is attributed to the concentrated power each user receives from the narrow, collimated, near-circular beams transmitted by the VCSEL transmitters. Further to that, VCSELs consume less power than other laser-light emitting devices as a result of the placement of Distributed Bragg Reflectors (DBRs), which lower the threshold current required to achieve propagation.

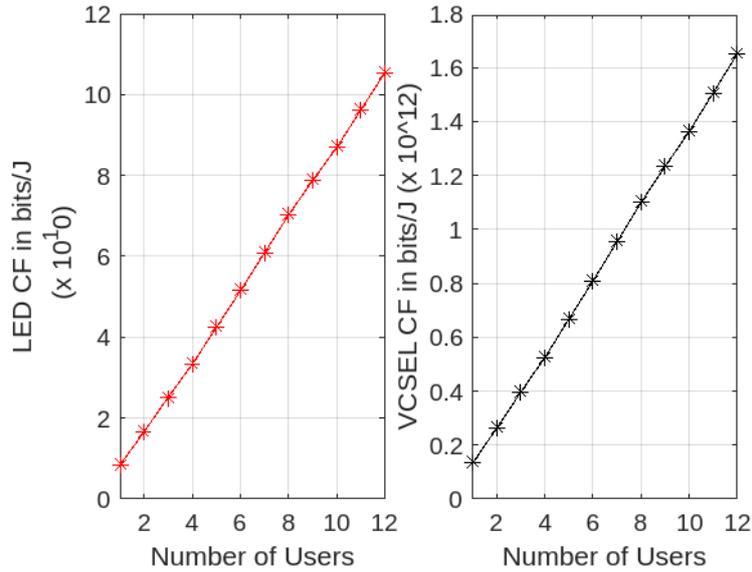

*Figure 3: Consumption Factor vs Number of Users.*

## 4. CONCLUSION

In this paper, we evaluated the energy efficiency of a laser-based OWC system, which uses VCSELs to transmit power through narrow, collimated, near-circular beams with low divergence angles to receivers. We compare the performance of the VCSEL system to an LED system, using equivalent electrical components and system models for a fair comparison. Our system model comprises multiple APs serving multiple receivers distributed randomly on the communication floor. ZF is implemented to support multiple access transmission. Our results show that the VCSEL system performs better than the LED system owing to the high bandwidth, the transmit beam profile, which allows the receivers to collect most of the transmitted power, and the lower threshold current of the VCSEL.

**ACKNOWLEDGEMENTS**

This work has been supported by the Engineering and Physical Sciences Research Council (EPSRC), in part by the INTERNET project under Grant EP/H040536/1, and in part by the STAR project under Grant EP/K016873/1 and in part by the TOWS project under Grant EP/S016570/1. All data are provided in full in the results section of this paper.